# Skyglow inside your eyes: intraocular scattering and artificial brightness of the night sky

Salvador Bará[1,*], Carmen Bao-Varela[2]

[1] Agrupación Astronómica Coruñesa Ío, 15005 A Coruña, Galicia
[2] Photonics4life Group, Facultade de Física e Facultade de Óptica e Optometría, iMATUS, Universidade de Santiago de Compostela. Galicia (Spain)

**Abstract**

The visual perception of the natural night sky in many places of the world is strongly disturbed by anthropogenic light. Part of this artificial light is scattered in the atmosphere and propagates towards the observer, adding to the natural brightness and producing a light polluted sky. However, atmospheric scattering is not the only mechanism contributing to increase the visual skyglow. The rich and diverse biological media forming the human eye also scatter light very efficiently and contribute, in some cases to a big extent, to the total sky brightness detected by the retinal photoreceptors. In this paper we quantify this effect and assess its relevance when the eye pupil is illuminated by light sources within the visual field. Our results show that intraocular scattering constitutes a significant part of the perceived sky brightness at short distances from streetlights. These results provide quantitative support to the everyday experience that substantial gains in naked-eye star limiting magnitudes can be achieved by blocking the direct light from the lamps that reaches the eye pupil. Urban lighting designs that reduce the illuminance over the eye pupil and locate the sources at large angles with respect to the visual axis are expected to help decreasing the artificial intraocular skyglow.

*Keywords: sustainable lighting, scattering, light pollution, sky brightness, radiometry, eye.*

## 1. Introduction

Significant progress has been made in the last two decades in modelling and quantifying the artificial brightness of the night sky. One of the physical processes playing a central role in the buildup of artificial skyglow is elastic scattering, the interaction of light with matter in which the former changes its propagation direction without changing its frequency. Atmospheric scattering from artificial light emissions has been studied by several research groups and practicable models for calculating its effects on skyglow are nowadays available [1-13]. Recent advances in this field include models of atmospheric scattering at short distances from the sources (from zero to about a few hundred meters), which provide interesting information about the skyglow components in urbanized spaces [14, 15]. Admittedly, many open questions still remain, concerning, e.g., the appropriate quantification of artificial light emissions, the characterization of the state of the atmosphere at every particular moment, and the optimum ways of performing the required calculations with sufficient accuracy and precision in a reasonable computing time.

But atmospheric scattering is not the only cause of artificial skyglow: the light scattered within the eye also plays an important role in the visual appearance of the sky, especially when the observer is close to bright artificial sources. The human eye contains a rich distribution of scatterers of different sizes, resulting from small scale inhomogeneities of the tissues and fluids composing the ocular media (including the cornea, aqueous humor, iris, eye lens, vitreous body, sclera, and the different retinal layers). Whereas most of the light reaching the eye pupil from a pointlike source is focused onto a small image spot on the retina, a fainter but non-negligible fraction of light is spread across a much larger retinal area, potentially giving rise to glare.

The optics of physiologically healthy eyes is far from being perfect, as the German anatomist and physicist Hermann von Helmholtz (1821–1894) cleverly pointed out [16]. Indeed, the bright core of the retinal image of a point source is neither a mathematical point, as geometrical optics would predict, nor a perfectly defined Airy disc, as elementary diffraction theory would suggest. Eye ametropies and higher-order geometrical eye aberrations cause this image to have an irregular shape, called the aberrated point spread function (PSF) of the eye [17]. This PSF typically subtends several arc-minutes of visual angle and its detailed features are highly variable across the population [18]. In spite of this variability, the PSFs of many healthy eyes tend to show some variation of a 'starlike' shape [19, 20], lending physical support to the usual depiction of bright stars as sources of rays, a common iconographic representation of celestial bodies across many cultures of the world.

*S. Bará, E-mail address: salva.bara@usc.gal

The intraocular scatter light pattern is considerably less intense than the PSF core but is much wider, extending across several tens of degrees in the retina [21, 22]. This scattered light is easier to perceive when an intense, pointlike source is viewed against a dark background as it happens when streetlights or vehicle lights are gazed outdoors at night [23]. The scattered light adds to the retinal illuminance, reducing the luminance constrast of the objects located across the visual field and potentially giving rise to disability glare. During the day or in illuminated indoors spaces at night the intraocular scattering is of course still present, but tends to be much less conspicuous than outdoors at nighttime: under the former circumstances the brightness in any direction of the visual field is clearly dominated by the direct luminance of the objects located in that direction, and the scattered light from the remaining points of the field does not lead to a significant decrease of visual performance for everyday life tasks, excepting in cases of strong scattering, as e.g., in eyes with cataracts [21].

In this paper we quantify the effect of the intraocular scattering on skyglow and compare it with other terms of artificial brightness of the night sky. The standard general equation relating the equivalent glare luminance to the pupil illuminance produced by a glaring source is briefly revisited in section 2. Section 3 applies this result to evaluate the relative contribution of intraocular scattering to the total skyglow in a nominal urban setting. Discussion and conclusions are drawn in sections 4 and 5, respectively.

## 2. The general glare equation

The standard description of intraocular scattering is based on the concept of equivalent luminance, $L_{eq}$, that is, the luminance of an external light field that would produce the same visual effect as the light actually scattered within the eye. The quantitative evaluation of this luminance is made with the help of the glare point-spread function [21, 22], $\Psi(\theta)$, defined as

$$\Psi(\theta) = L_{eq}(\theta)/E_{gl} \ , \tag{1}$$

where $E_{gl}$ is the illuminance produced on the input pupil of the eye by a pointlike glaring source, and $\theta$ is the visual angle between the source and the line of sight (Figure 1). The scattering PSF has units of inverse steradians (sr$^{-1}$) and is normalized such that its integral across the visual field is unity.

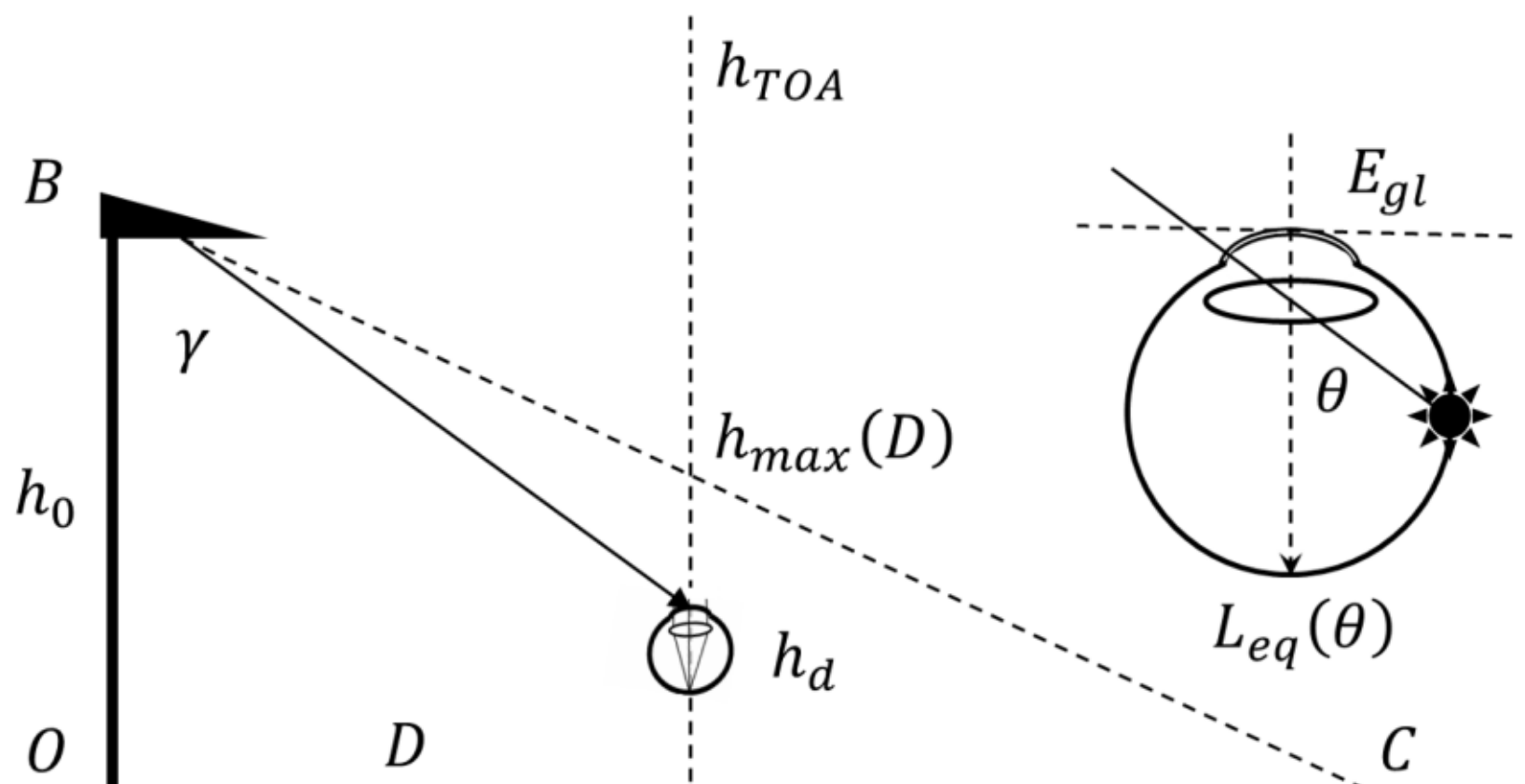

**Fig. 1.** Geometry for evaluating the equivalent sky brightness (glare luminance, $L_{eq}(\theta)$) produced by a streetlight at height $h_0$ located at $O$, on an observer of height $h_d$ located a distance $D$ away. The scattering angle $\theta$ is here depicted for zenith observation, hence $\theta = \gamma$. The line $BC$ marks the upper limit of the air column illuminated by the lamp, whose height above ground $h_{max}$ depends on the distance to the streetlight. The atmosphere is assumed to be extended up to an altitude $h_{TOA}$. For the interpretation of the schematic eye, see section 4 Discussion.

Several expressions with different degrees of complexity have been proposed for $\Psi(\theta)$. For the purposes of the present work, we will use the so-called *general glare equation* [21], which is valid for $\theta$ ranging from 0.1° to the limit of the visual field (~100°):

$$\Psi(\theta) = \frac{10}{\theta^3} + \left(\frac{5}{\theta^2} + \frac{0.1 \cdot p}{\theta}\right)\left[1 + \left(\frac{A}{62.5}\right)^4\right] + 2.5 \cdot 10^{-3} \cdot p \ , \tag{2}$$

where $\theta$ is expressed in degrees, $A$ is the age of the observer (in years) and $p$ is an iris pigmentation factor, equal to 0 for dark eyes, 0.5 for brown, and 1.0 for blue-green ones.

The following figures show the value of $L_{eq}(\theta)$ for different combinations of $E_{gl}$, $A$ and $p$. In standard SI units the corneal illuminance $E_{gl}$ is expressed in lux (lx), and the value of $L_{eq}$ is consequently obtained in candela per square meter, $cd \cdot m^{-2}$ (recall that the intraocular scattering PSF has units $sr^{-1}$). For facilitating the comparison with other skyglow terms, the values of $L_{eq}$ will be expressed here in an astronomical scale of magnitudes per square arcsecond, $mag/arcsec^2$. The CIE photopic spectral luminous efficiency function [24] is the native photometric band for defining the astronomical magnitude system corresponding to the brightness in $cd \cdot m^{-2}$ [25]. However, the Johnson-Cousins V band [26] can also be used for this purpose. This facilitates the comparisons with other works, since skyglow contributions are frequently expressed or measured in this astronomical band. It shall be kept in mind that the accurate conversion from $cd \cdot m^{-2}$ to $mag_V/arcsec^2$ depends on the spectrum of the light [27, 28]. For the figures below we have chosen a transformation based on an equivalence of 200 $\mu cd \cdot m^{-2}$ to 22.0 $mag_V/arcsec^2$, that is, a zero-point luminance $L_0 = 1.26 \cdot 10^5$ $cd \cdot m^{-2}$. This corresponds to light fields of correlated color temperatures (CCT) close to 3000 K, not untypical of urban lights and light polluted skies (cfr. definitions and Fig 3(a) in [27]). The transformation between these two systems of units is then

$$L_{eq}(\mathrm{mag_V/arcsec^2}) = -2.5 \ \log_{10}\left[\frac{L_{eq}(\mathrm{cd \cdot m^{-2}})}{1.26 \cdot 10^5}\right] \tag{3}$$

The curves in Fig. 2 show the values of $L_{eq}$ across the whole range of scattering angles, for pupil illuminances between 0.01 and 100 lx. These curves were calculated for an observer of age $A$=35 yr and eye pigmentation factor $p$=0.5.

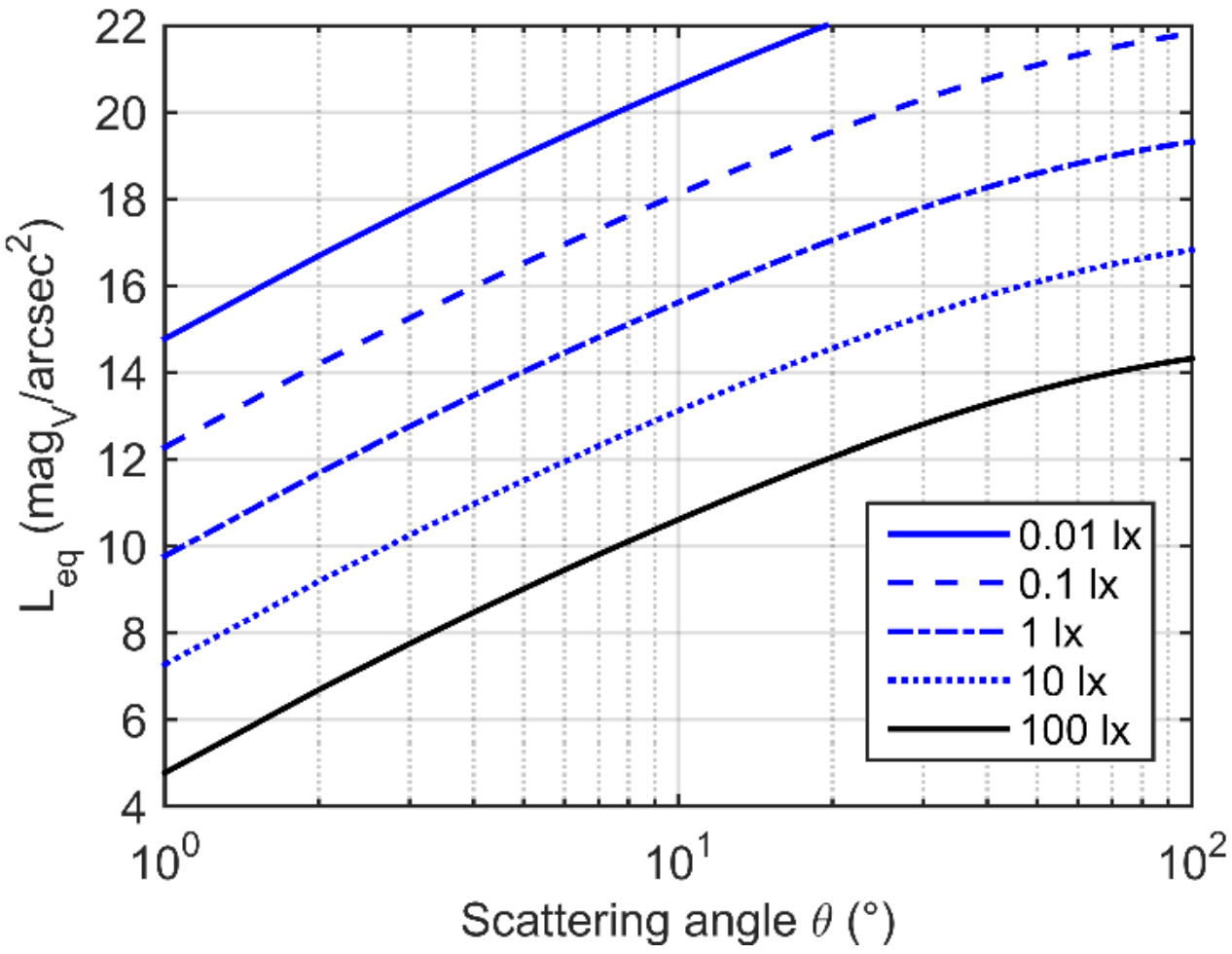

**Fig. 2.** Value of $L_{eq}$, the equivalent sky brightness (glare luminance), for several amounts of illuminance on the eye pupil (in lx). Average observer of age 35 yr and pigmentation factor $p$=0.5 When looking toward the zenith, the relevant input is the horizontal illuminance on the pupil of the eye. In urban settings values of order of several tens of lx are not unusual. Note that 10° away from the image of the streetlight the scattered light may produce an artificial sky luminance of about 12 to 16 $mag_V/arcsec^2$, considerably larger than the typical sky brightnesses due to atmospheric scattering.

Figure 3 shows the dependence of the equivalent luminance $L_{eq}$ on the age of the average observer, for $A$= 20, 40, 60, and 80 years, and eyes with a medium pigmentation factor, $p$=0.5, under 1 lx pupil illumination. It is well known that intraocular scattering tends to increase with age, mostly due to structural changes in the eye lens (crystalline) [21, 29]. This amounts to an increase of brightness of order –1 $mag_V/arcsec^2$ for 80 yr old observers across the whole range of scattering angles, in comparison with young 20 yr old observers.

Finally, in Fig.4 we show the behavior of $L_{eq}$ for different levels of eye pigmentation, assuming an observer of 40 yr of age and a pupil illuminance of 1 lx. The differences caused by pigmentation tend to be more noticeable at mid- to large scattering angles, in the region of $\theta$ above ~20°.

Recall that the luminance values shown in these figures are exclusively due to the light scattered within the eye. In an actual urban setting, additional terms of light either directly or after scattering in the atmosphere also contribute to the total luminance perceived by an observer. Calculating these contributions and comparing the intraocular one with them is the topic of the next section.

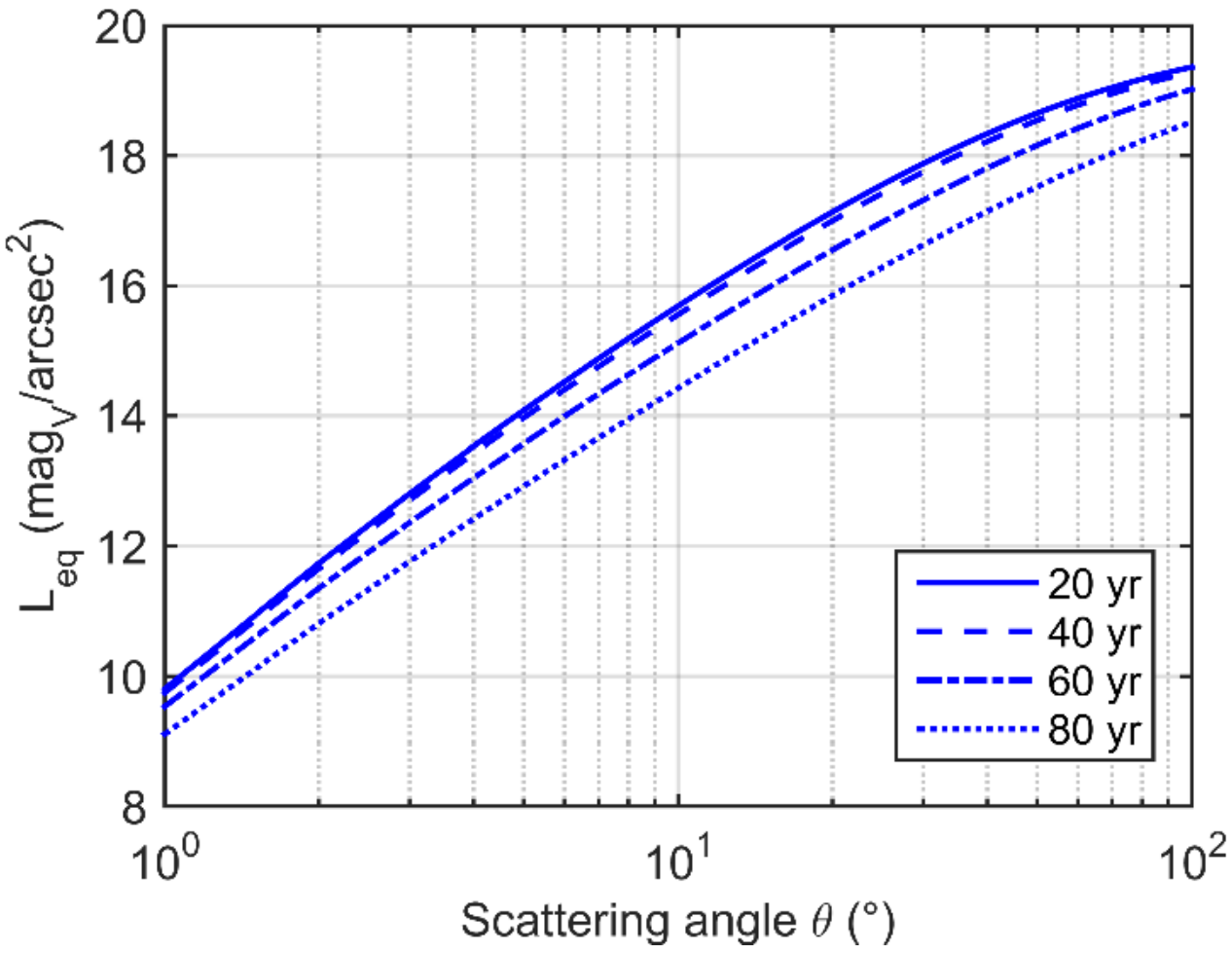

**Fig. 3.** Value of $L_{eq}$ for average observers of ages 20, 40, 60, and 80 yr. Pupil illuminance was set to 1 lx, in eyes of medium pigmentation, *p*=0.5. An increase of brightness of order –1 $mag_V/arcsec^2$ is produced across the whole range of scattering angles for the oldest observers in comparison with the youngest ones.

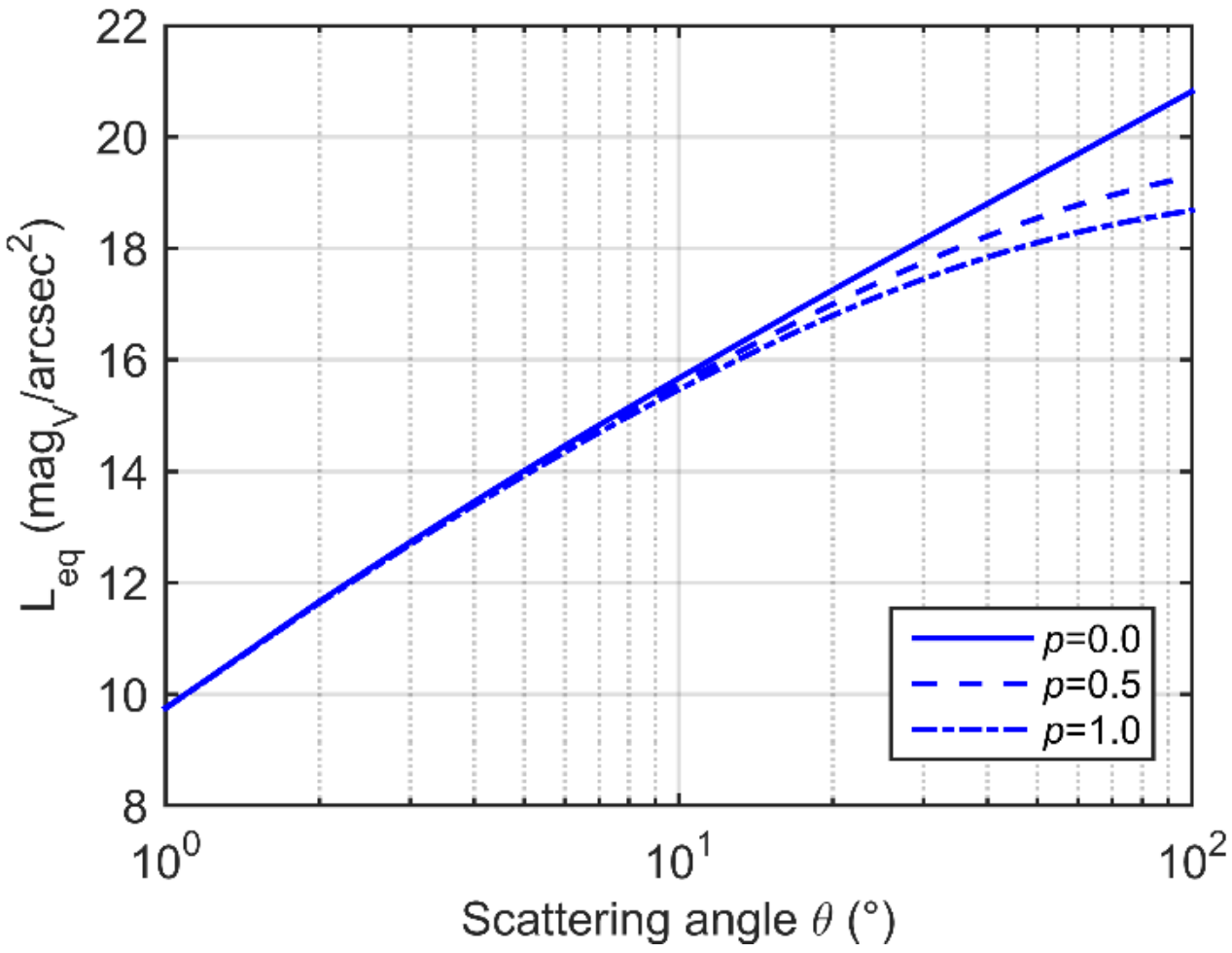

**Fig. 4.** Value of $L_{eq}$ for an average observer of age 40 yr and different pigmentation factors (*p*=0.0, 0.5, 1.0) under pupil illuminance of 1 lx. Differences due to pigmentation are more noticeable at scattering angles larger than ~20°.

## 3 How does intraocular scattering compare with the remaining skyglow terms?

The relative weight of the different terms of artificial skyglow is of course contingent on the detailed configuration of the illuminated space surrounding the observer. To answer the question that motivates this paper it is convenient to analyze a simplified situation that nevertheless captures the basic physics of the problem. In the following subsections a situation of this kind is described, with the aim of obtaining some useful insights about the relative importance of the different skyglow contributions in a typical urban nightscape.

### *3.1. Model and parameters*

Let us consider an observer gazing up at the zenith (Fig. 1), from a point at a horizontal distance *D* from a luminaire *OB*. The lamp *B*, at height $h_0$, emits light within a limited range of angles, being the line *BC* the upper boundary of the air column directly illuminated by its radiance. The observer eyes are at height $h_d$ above ground. The angle $\gamma$ is the nadir angle of the eye as seen from the lamp. For zenith gazing this angle is equal to $\theta$, the angle of incidence of the direct lamp radiance on the eye pupil plane.

In such situation, the sky brightness may be considered the result of the contribution of four main terms: (a) the baseline brightness of the natural sky, (b) the skyglow produced by the ensemble of city sources, excluding the analyzed streetlight $B$, (c) the radiance from $B$ scattered in the air column above the observer, from $h=h_d$ to $h=h_{max}(D)$, and (d) the radiance from $B$ scattered within the observer's eye.

As a cautionary note, recall that the intraocular scattering is produced by all sources of light present within the field of view of the eye. This means that the natural sky is also a contributor: each individual star acts as a pointlike source, as well as each diffuse square arcsecond patch of the sky and, more conspicuously, the Moon and bright planets. The same applies to the remaining city lights, apart from the one here individualized ($B$): their radiances, both the direct and the scattered in the atmosphere, act as sources of intraocular scattered light. In the following subsections we will restrict our analysis to the intraocular scattered light due only to $B$, given that at close distances this lamp will be the dominant contributor in moonless nights. Note however that there would be no fundamental difficulty in extending the intraocular scattering analysis to encompass a constellation of neighboring glaring sources, including the diffuse brightness of façades and vertical walls, and also the sky (see section 4, Discussion).

#### *3.1.1. Dark sky brightness*

The nighttime brightness of the natural sky, for any photometric band and given field-of-view, is a highly variable physical magnitude. In moonless nights, after the astronomical twilight, this brightness depends on the celestial objects within the field of view including individual stars and planets, diffuse astrophysical sources (zodiacal light, background galactic and extragalactic light), and terrestrial sources (airglow). Most of the light from these sources arrives directly to the eye/detector, but a non-negligible part is detected after undergoing one or more scattering events. The state of the atmosphere determines the amount of radiation that is attenuated (scattered or absorbed). A comprehensive model for the brightness of the natural sky in several photometric bands, including the effects of first-order atmospheric scattering, can be found in [30, 31]. The zenith brightness in the Johnson-Cousins V band under stable atmospheric conditions is highly variable across the year, and largely dependent on the latitude of the observer (which determines what celestial regions can be seen at the zenith) [30]. For the purposes of the present work, we use as a reference brightness of the dark sky the conventional value $m_{DS} = 22.00$ $\mathrm{mag_V/arcsec^2}$.

#### *3.1.2. Skyglow due to the surrounding city sources*

The general procedure for evaluating the contribution of the ensemble of sources surrounding the observer is described in [8]. For the present example we have used the PSF corresponding to the emission of flat horizontal sources (illuminated pavements) described in section 3.2 of [14], assuming a city of radius 5 km with average emissions of 4.0 $\mathrm{Mlm/km^2}$ towards the pavements. For a city with half of its surface occupied by buildings this corresponds to an average illumination of 8.0 lx on the illuminated spaces (streets, roadways and other illuminated outdoor areas). Pavement reflections were assumed Lambertian, and the average reflectance of the city surfaces in the CIE photopic band was set to 0.2. The eyes of the observer are located at 1.70 m above ground. The atmospheric parameters are the same as those used in section 3.1. of reference [14]. As above, the zero-point luminance $L_0 = 1.26\cdot 10^5$ $\mathrm{cd\cdot m^{-2}}$ has been adopted to relate the Johnson V magnitudes per square arcsecond to the luminance. The estimated skyglow contribution of the city under these conditions is $m_{City} = 17.83$ $\mathrm{mag_V/arcsec^2}$.

#### *3.1.3. Radiance from the streetlight B scattered in the air column above the observer*

The procedure for the calculation of this radiance is described in section 3.3. of [14]. For our present example the lamp height above ground has been set to 8 m, its nadir emission angles span the range $\gamma \in [0°, 75°]$, its active surface has an area $A_s = 0.01$ $\mathrm{m^2}$ (10x10 $\mathrm{cm^2}$ emitting board) and the lamp luminance $L_s$ is constant within this angle range, scaled such that the average luminous flux on the ground per square meter equals the average illuminance value of the city streets (8.0 lx, see above). The resulting luminance at the eyes of the observer is transformed to Johnson V magnitudes per square arcsecond using the same zero-point luminance as above, obtaining that way the contribution $m_{str}$ which, of course, depends on $D$.

#### *3.1.4. Intraocular scattering*

The geometry of observation in our present example implies that for every distance $D$ of the observer to the lamp, the nadir angle $\gamma$ at which the lamp radiance $L_s$ is emitted towards the eye pupil equals the scattering angle for foveal vision of the zenith, $\theta$. The expression of the glaring illuminance over the eye pupil is then $E_{gl} = L_s\, A_s \cos^2\theta / [(h_0 - h_d)^2 + D^2]$, from which the equivalent luminance $L_{eq}(\theta)$ can be calculated using Eqs (1)- (2). The result can be transformed into Johnson V magnitudes per square arcsecond, $m_{ioc}$, using Eq. (3). We have considered an observer 40 yr old with medium eye pigmentation, $p$=0.5.

The total luminance is the sum of the four luminance contributions described in the above subsections, expressed in cd·m$^{-2}$. Note that whereas the luminances in cd·m$^{-2}$ are additive, the luminances expressed in mag$_V$/arcsec$^2$ are not. If the magnitudes per square arcsecond $\{m_i\}$ associated with $i$=1,...,$N$ luminance terms are all expressed in the same magnitude system (i.e., same photometric band, same measured variable, and same zero point, which is our case here), the resulting magnitude per square arcsecond $m$ associated with the total luminance $L_{tot}$ is given by:

$$m = -2.5 \log_{10}\left[\sum_{i=1}^{N} 10^{-0.4 \cdot m_i}\right] \quad (4)$$

### *3.2. Results*

The results of the above calculations, as a function of the horizontal distance $D$ between the observer and the streetlight, are summarized in Fig. 5. The individual skyglow contributions are displayed with red lines, whereas the total luminance is shown in blue with square symbols. The vertical axis has units mag$_V$/arcsec$^2$, and the values corresponding to the total luminance have been calculated combining the individual magnitudes of the four contributions using Eq. (4).

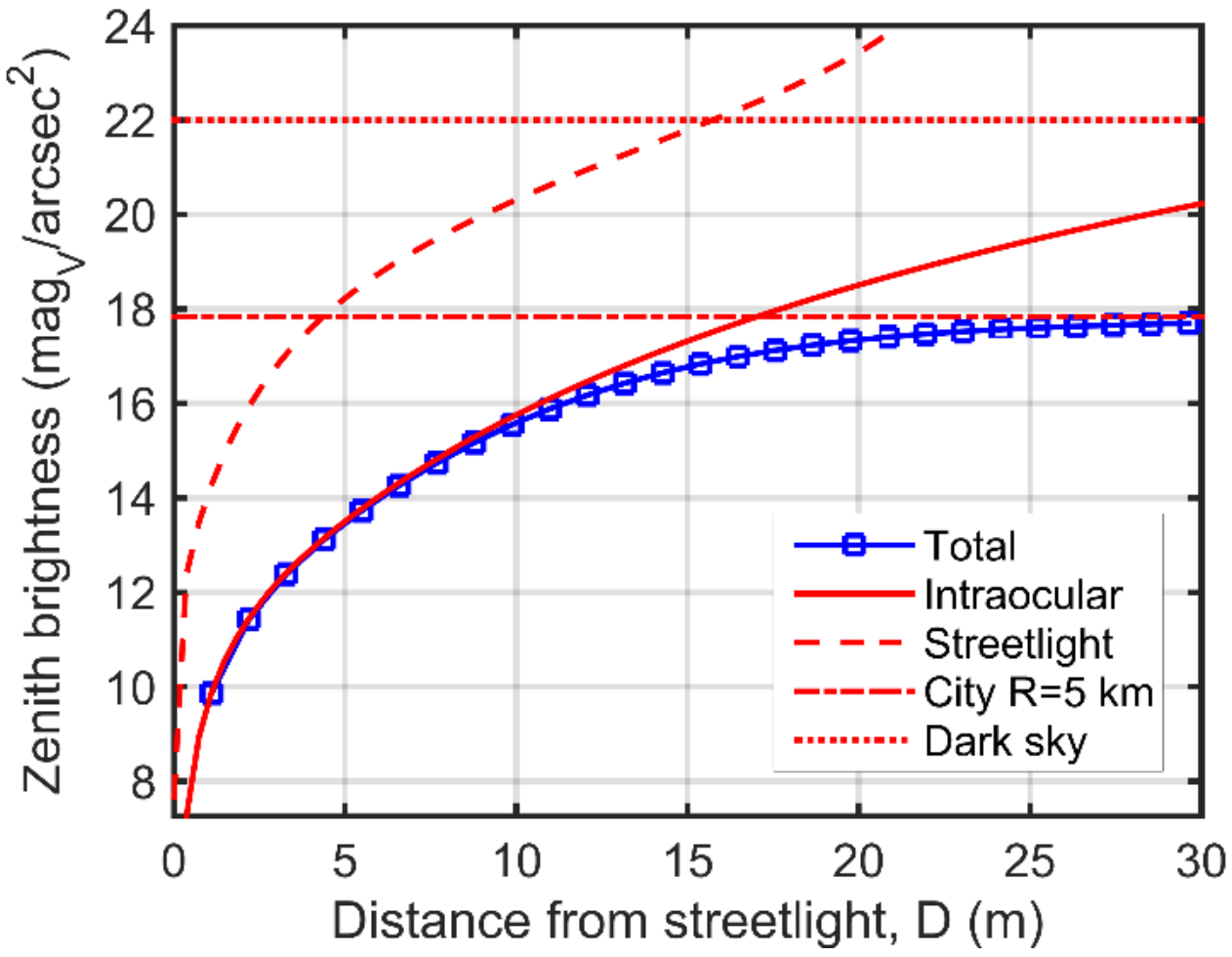

**Fig. 5.** Total zenith brightness and its four basic components included in this example, expressed in mag$_V$/arcsec$^2$. The transformation from luminances in cd·m$^{-2}$ to Johnson V magnitudes per square arcsecond is made using the zero-point luminance $L_0$ =1.26·10$^5$ cd·m$^{-2}$. See text for details.

According to these results, valid for the example here analyzed, the zenith brightness is dominated by the intraocular scattering when the observer is located at distances shorter than ~17 m from the streetlight, point at which the intraocular scattering has the same strength as the surrounding city contribution. At larger distances the total brightness steadily approaches the sum (in cd·m$^{-2}$) of the city and the dark sky contributions. The luminance produced by the streetlight scattering in the air column above the observer, spanning the height range $h$=$h_d$ to $h$=$h_{max}(D)$, decreases with the distance $D$ and becomes zero (in cd·m$^{-2}$) when the upper boundary of the air column directly illuminated by the lamp falls below the height of the eyes of the observer, $h_d$ =$h_{max}(D)$, something that in our example happens at $D$=23.51 m (from that point on, the magnitudes per square arcsecond associated to this term are +∞). It is worth noting that within the first few meters (~4 m) from the lamp this scattered luminance is larger than the one due to the rest of the city, but it is still about +4 mag$_V$/arcsec$^2$ fainter than the intraocular term (and this difference increases for larger distances).

## 4 Discussion

This paper describes how the intraocular scattering increases the visual brightness of the night sky, when the observer is located close to a streetlight. The numerical results for a particular but not unrealistic example show that the intraocular contribution to the zenith sky brightness is clearly dominant at short to medium distances from the lamppost, steadily converging to the combined city and natural sky background when this distance increases. Substantial gains in the quality of the night sky can be expected by blocking the direct radiance entering the eye from the nearby lamp. For the example in section 3 these gains amount to an increase in darkness of almost +5 $\mathrm{mag_V/arcsec^2}$ at 4 m from the lamppost and about +1 $\mathrm{mag_V/arcsec^2}$ at 15 m from it. These quantitative calculations provide numeric support to the common experience of perceiving a richer starry sky when the direct light from the lamp is blocked by the observers by e.g. using their hands as lateral eye screens. Transitioning from +13 to +18 $\mathrm{mag_V/arcsec^2}$ the naked-eye limiting magnitude increases from –0.69 to +3.97 [32]. Note however that the resulting sky quality will still be limited by the large contribution of the surrounding city.

We have used a classical, standard model for calculating the equivalent luminance of the intraocular scattering, $L_{eq}$, when the eye pupil is illuminated by a glaring illuminance, $E_{gl}$, from a pointlike source. Both quantities are related through the intraocular scattering PSF, $\Psi(\theta)$, which describes the spread of the scattered light across the observer's retina in terms of the scattering angle $\theta$, which is the visual angle between the direction of the glaring source (assumed pointlike) and the line of sight. The standard PSFs represent the scattering for an average observer, with controls of age and eye pigmentation. The PSF model used in this paper, Eq. (2), is valid with enough accuracy for scattering angles within the interval $\theta \in [0.1°, 100°]$, which is more than enough for our present purposes. Other standard forms of the PSF, with wider or narrower angular ranges of validity, may be found in [21, 22]. Note that the structure of the retinal image at smaller angles (i.e. $\theta \leq 6$ arcmin) is strongly determined by the particular features of the geometric and diffractive eye aberration pattern of each person [17-20], rather than by the diffuse scattering term.

The reader might have noticed that the schematic eye model in the inset of Fig.1 depicts the light ray path as if there were no refraction in the ocular media, which is of course an oversimplification; the light rays do change their propagation direction when the refractive index has a non-null gradient, something that happens at the interfaces of different optical media (e.g. the discontinuities in the air-tearfilm-cornea boundaries) or within them (e.g. the graded-index structure of the crystalline lens). The rigorous display of $\theta$ in the object and retinal image spaces should be made using the eye nodal points as angle vertices, but this optical formality has been omitted in that figure for the sake of simplicity. Also, in this figure the pupillary axis and the visual axis (line of sight) have been made to coincide, whereas actually there is a small angle between them, mostly due to the slightly eccentric location of the eye fovea. We have addressed in this work the total intraocular scattering, considering the eye optical system as a whole. A detailed description of the relative contribution of each eye element to the global scattering both in physiologically normal eyes and in some pathological conditions can be found in van den Berg's works [21,22].

There is in principle no difficulty for evaluating the effects of the intraocular scattering for all sources whose light enters the eye, not only for the nearest streetlight as done in this work. To do so, recall that the total luminance $L(\boldsymbol{\alpha})$ perceived in any visual direction described by the unit vector $\boldsymbol{\alpha}(z,\varphi)$, where $z$ is the polar angle from the line of sight and $\varphi$ is the azimuth, both measured in the eye reference frame, is the sum of the direct luminance entering the eye from that direction, $L_f(\boldsymbol{\alpha})$, plus the all the light scattered within the eye in that particular direction, $L_{ioc}(\boldsymbol{\alpha})$, due to the sources present within the field, $L_s(\boldsymbol{\alpha}')$. The irradiance on the eye pupil produced by a small patch of the external light field centered in $\boldsymbol{\alpha}'(z',\varphi')$ and of elementary solid angle extent $\mathrm{d}\Omega = \sin z' \mathrm{d}z' \mathrm{d}\varphi'$ (units sr) is $\mathrm{d}E(\boldsymbol{\alpha}') = L_f(\boldsymbol{\alpha}') \cos z' \,\mathrm{d}\Omega$. According to Eq.(1) this illuminance will produce an intraocular scattered light distribution whose value in the direction $\boldsymbol{\alpha}$ will be given by $\mathrm{d}L_{eq}(\boldsymbol{\alpha},\boldsymbol{\alpha}') = \Psi(\theta)\, L_f(\boldsymbol{\alpha}') \cos z' \,\mathrm{d}\Omega$, where $\theta$, the scattering angle, is the angle formed by the unit vectors $\boldsymbol{\alpha}$ and $\boldsymbol{\alpha}'$, that is $\theta = \arccos(\boldsymbol{\alpha}\cdot\boldsymbol{\alpha}')$, where the dot stands for scalar product. Hence the total radiance perceived in the direction $\boldsymbol{\alpha}$ will be:

$$L(\boldsymbol{\alpha}) = L_f(\boldsymbol{\alpha}) + \int_\Omega \Psi[\arccos(\boldsymbol{\alpha}\cdot\boldsymbol{\alpha}')] L_f(\boldsymbol{\alpha}') \cos z' \sin z' \mathrm{d}z' \mathrm{d}\varphi', \quad (5)$$

where the integral is extended to $\Omega$, the region of the space of directions (approximately hemispheric) from which the light enters directly the eye. Note that for the correct application of this equation the PSF $\Psi(\theta)$ should be the 'full model' one, that is, defined for $\theta$ starting from 0° [21, 22].

The results presented in this work are affected by several limitations. On the one hand, the inter-subject variability is not captured by the average PSF, $\Psi(\theta)$, hence the relative importance of the intraocular scattering term may vary depending on the observer. We also have not dealt with the chromatic aspects of the intraocular scattering field, for instance with the color structure of the scattering 'needles' [33], a phenomenon easy to perceive when the luminance of the lamps is very high. We have also restricted to the first scattering order our modeling of the light scattered in the air column, emitted by the streetlight under study and by the surrounding city. This approximation is arguably valid for the nearby streetlight, given the short distances involved and the usual optical depth of the atmosphere in stargazing nights, albeit it may be not so in case of thick fog or for the city lights located several tens of km away. Also, we have used the standard value of 22.00 $\mathrm{mag_V/arcsec^2}$ for the natural sky darkness in the Johnson V band, which should be considered only as an approximate relative baseline: the night sky has a great variability depending on the celestial objects located within the field of view. More importantly, we have not included the Moon in the calculations of Section 3. Both the direct light of the Moon, the moonlight scattered in the atmosphere and the one scattered within the eye may give rise to remarkable increases of the visual brightness of the night sky.

## 5 Conclusions

Intraocular scattering is an integral part of the visual sky brightness, and it may be the dominant contribution when the observer is located close to high luminance streetlights. In this work we describe the basic equations for quantifying this effect and compare it with the remaining artificial skyglow terms. The results provide quantitative support for the common experience of improving the perceived night sky quality by blocking the direct radiance from the streetlights before it reaches the observer pupil. Depending on the characteristics of the sources, the observing geometry and the observer personal features, the naked-eye limiting magnitude may increase in a few units by means of this simple procedure. The resulting sky quality, however, will still be limited by the overall emissions of the city and surrounding territories.

## Acknowledgements

CB-V acknowledges funding from Xunta de Galicia/FEDER, grant ED431B 2020/29

## Declaration of Competing Interest

The authors declare that they have no known competing financial interests or personal relationships that could have appeared to influence the work reported in this paper.

## References

[1] Garstang, R. H. (1989). Night-sky brightness at observatories and sites. PASP 101, 306. doi:10.1086/132436

[2] Cinzano, P., & Falchi, F. (2012).The propagation of light pollution in the atmosphere. Monthly Notices of the Royal Astronomical Society 427(4), 3337–3357. doi: 10.1111/j.1365-2966.2012.21884.x

[3] Falchi, F., Cinzano, P., Duriscoe, D., Kyba, C. C. M., Elvidge, C. D., Baugh, K., Portnov, B. A., Rybnikova, N. A., & Furgoni, R. (2016). The new world atlas of artificial night sky brightness. Sci. Adv. 2, e1600377, doi: 10.1126/sciadv.1600377

[4] Kocifaj, M. (2007). Light-pollution model for cloudy and cloudless night skies with ground-based light sources. Applied Optics 46(15), 3013-22. doi: 10.1364/AO.46.003013

[5] Kocifaj, M. (2018). Multiple scattering contribution to the diffuse light of a night sky: A model which embraces all orders of scattering. Journal of Quantitative Spectroscopy and Radiative Transfer, 206, 260–72. doi: 10.1016/j.jqsrt.2017.11.020

[6] Bará, S., & Lima, R. C. (2018). Photons without borders: quantifying light pollution transfer between territories, International Journal of Sustainable Lighting 20(2), 51-61. doi:10.26607/ijsl.v20i2.87

[7] Bará, S., Rigueiro, I., & Lima, R. C. (2019). Monitoring transition: Expected night sky brightness trends in different photometric bands. Journal of Quantitative Spectroscopy and Radiative Transfer, 239, 106644. doi: 10.1016/j.jqsrt.2019.106644

[8] Bará, S., Falchi, F., Lima, R. C. & Pawley M. (2021). Can we illuminate our cities and (still) see the stars?. International Journal of Sustainable Lighting 23(2), 58-69. doi: 10.26607/ijsl.v23i2.113

[9] Falchi, F., & Bará, S. (2021). Computing light pollution indicators for environmental assessment. Natural Sciences, e10019. doi: 10.1002/ntls.10019

[10] Duriscoe, D. M., Anderson, S. J., Luginbuhl, C. B., & Baugh, K. E. (2018). A simplified model of all-sky artificial sky glow derived from VIIRS Day/Night band data. Journal of Quantitative Spectroscopy and Radiative Transfer, 214, 133–145. doi: 10.1016/j.jqsrt.2018.04.028

[11] Aubé, M., & Simoneau, A. (2018). New features to the night sky radiance model illumina: Hyperspectral support, improved obstacles and cloud reflection. Journal of Quantitative Spectroscopy and Radiative Transfer, 211, 25–34. doi: 10.1016/j.jqsrt.2018.02.033

[12] Simoneau, A., Aubé, M., Leblanc, J., Boucher, R., Roby, J., & Lacharité, F. (2021). Point spread functions for mapping artificial night sky luminance over large territories, Monthly Notices of the Royal Astronomical Society, 504(1), 951–963. doi: 10.1093/mnras/stab681

[13] Linares, H., Masana, E., Ribas, S. J., Aubé, M., Simoneau, A., & Bará, S. (2020). Night sky brightness simulation over Montsec protected area. Journal of Quantitative Spectroscopy and Radiative Transfer, 249, 106990. doi: 10.1016/j.jqsrt.2020.106990

[14] Bará, S., Bao-Varela, C. & Kocifaj, M. (2022). Modelling the artificial night sky brightness at short distances from streetlights. Journal of Quantitative Spectroscopy and Radiative Transfer (in press), doi: 10.1016/j.jqsrt.2022.108456

[15] Zamorano, J., Bará, S., Barco, M., García, C., & Caballero, A.L. (2023). Controlling the artificial radiance of the night sky: the Añora urban laboratory. Journal of Quantitative Spectroscopy and Radiative Transfer 296, 108454. doi: 10.1016/j.jqsrt.2022.108454

[16] Biedermann, K. (2002). "The Eye, Hartmann, Shack, and Scheiner," in Optical Information Processing: A Tribute to Adolf Lohmann, H. John Caulfield, Editor; Chapter 7, pp 123 - 130. SPIE PRESS, Bellingham, Washington, USA.

[17] Navarro R., Moreno, E. & Dorronsoro C. (1998). Monochromatic aberrations and point-spread functions of the human eye across the visual field, Journal of the Optical Society of America A, 15, 2522-2529.

[18] Thibos, L.N., Hong, X., Bradley, A., & Cheng, X. (2002). Statistical variation of aberration structure and image quality in a normal population of healthy eyes. Journal of the Optical Society of America A, 19(12), 2329-2348.

[19] Navarro, R., & Losada, M. A. (1997). Shape of stars and optical quality of the human eye. Journal of the Optical Society of America A, 14, 353-359.

[20] Bará, S. (2013). The sky within your eyes (Eye aberrations and visual Astronomy). Sky and Telescope, 126 (2), 68-71.

[21] van den Berg, T. J. T. P., Franssen, L., & Coppens, J. E. (2010). Ocular Media Clarity and Straylight. In Darlene A. Dartt (ed), *Encyclopedia of the Eye*, Vol 3. Oxford:Academic Press; pp. 173-183

[22] van den Berg, T. J. T. P., Franssen, L., Kruijt, B., & Coppens, J. E. (2013). History of ocular straylight measurement: A review. Zeitschrift für Medizinische Physik, 23, 6–20. doi: 10.1016/j.zemedi.2012.10.009

[23] Bará S. (2014). Naked-eye Astronomy: Optics of the starry night skies, Proc. SPIE 9289, 12th Education and Training in Optics and Photonics Conference, 92892S. doi: 10.1117/12.2070764

[24] CIE, Commision Internationale de l'Éclairage. (1990). CIE 1988 2° SpectralLuminous Efficiency Function for Photopic Vision. CIE 86:1990. Vienna: Bureau Central de la CIE..

[25] Bará S. (2017). Variations on a classical theme: On the formal relationship between magnitudes per square arcsecond and luminance, International Journal of Sustainable Lighting 19(2), 104-111. doi: 10.26607/ijsl.v19i2.77

[26] Bessell, M. S. (1990). UBVRI passbands. Publications of the Astronomical Society of the Pacific, 102, 1181-1199. doi: 10.1086/132749

[27] Bará, S., Aubé, M., Barentine, J., & Zamorano, J. (2020). Magnitude to luminance conversions and visual brightness of the night sky. Monthly Notices of the Royal Astronomical Society, 493, 2429–2437. doi: 10.1093/mnras/staa323

[28] Fryc, I., Bará, S., Aubé, M., Barentine, J. C. & Zamorano, J. (2022). On the Relation between the Astronomical and Visual Photometric Systems in Specifying the Brightness of the Night Sky for Mesopically Adapted Observers, Leukos, 18:4, 447-458, doi:10.1080/15502724.2021.1921593

[29] van den Berg, T. J. T. P., van Rijn, L. J., Michael, R., et al. (2007). Straylight effects with aging and lens extraction. American Journal of Ophthalmology 144: 358–363. doi: 10.1016/j.ajo.2007.05.037.

[30] Masana, E., Carrasco, J. M., Bará, S., & Ribas, S. J. (2021). A multi-band map of the natural night sky brightness including Gaia and Hipparcos integrated starlight. Monthly Notices of the Royal Astronomical Society, 501, 5443–5456. doi 10.1093/mnras/staa4005

[31] Masana, E., Bará, S., Carrasco, J. M., & Ribas, S. J. (2022). An enhanced version of the Gaia map of the brightness of the natural sky. International Journal of Sustainable Lighting, 24(1), 1-12. doi:10.26607/ijsl.v24i1.119

[32] Schaefer, B. E. (1990). Telescopic limiting magnitudes, Publications of the Astronomical Society of the Pacific, 102, 212-229. Calculator online in http://unihedron.com/projects/darksky/NELM2BCalc.html

[33] van den Berg, T. J., Hagenouw M. P. , & Coppens, J. E. (2005). The ciliary corona: physical model and simulation of the fine needles radiating from point light sources. Investigative Ophthalmology & Visual Science, 46(7), 2627-32. doi: 10.1167/iovs.04-0935